\documentclass[aip, jap, amsmath,amssymb,reprint]{revtex4-1}

\usepackage{graphicx}
\usepackage{dcolumn}
\usepackage{bm}
\usepackage{color}

\newcommand{\av}[1]{\langle #1 \rangle}

\newcommand{\E}[1]{$\times$10$^{#1}$}
\newcommand{\mumax}{{MuMax$^3$} }
\newcommand{\m}{\ensuremath{\mathbf{m}}}
\newcommand{\Beff}{\ensuremath{\mathbf{B}_\mathrm{eff}}}

\begin{document}
\title{Adaptively time stepping the stochastic Landau-Lifshitz-Gilbert equation at nonzero temperature: implementation and validation in \mumax }
\author{J. Leliaert}
\email{jonathan.leliaert@ugent.be}
\affiliation{Department of Solid State Sciences, Ghent University, 9000 Gent, Belgium}
\author{J. Mulkers}
\affiliation{Department of Solid State Sciences, Ghent University, 9000 Gent, Belgium}
\affiliation{Department of Physics, Antwerp University, 2020 Antwerp, Belgium}
\author{J. De Clercq}
\affiliation{Department of Solid State Sciences, Ghent University, 9000 Gent, Belgium}
\author{A. Coene}
\affiliation{Department of Electrical Energy, Metals, Mechanical Constructions and Systems
, Ghent University, 9052 Zwijnaarde, Belgium}
\author{M. Dvornik}
\affiliation{Department of Physics, University of Gothenburg, 412 96, Gothenburg, Sweden}
\author{B. Van Waeyenberge}
\affiliation{Department of Solid State Sciences, Ghent University, 9000 Gent, Belgium}

\date{\today}

\begin{abstract}
Thermal fluctuations play an increasingly important role in micromagnetic research relevant for various biomedical and other technological applications. Until now, it was deemed necessary to use a time stepping algorithm with a fixed time step in order to perform micromagnetic simulations at nonzero temperatures. However, Berkov and Gorn\cite{BER-02a} have shown that the drift term which generally appears when solving stochastic differential equations can only influence the length of the magnetization. This quantity is however fixed in the case of the stochastic Landau-Lifshitz-Gilbert equation. In this paper, we exploit this fact to straightforwardly extend existing high order solvers with an adaptive time stepping algorithm. We implemented the presented methods in the freely available GPU-accelerated micromagnetic software package \mumax and used it to extensively validate the presented methods. Next to the advantage of having control over the error tolerance, we report a twenty fold speedup without a loss of accuracy, when using the presented methods as compared to the hereto best practice of using Heun's solver with a small fixed time step.
\end{abstract}
\maketitle

\section{Introduction}
\label{sec:introduction}
Micromagnetic simulations of systems at nonzero temperatures are an increasingly important tool to numerically investigate magnetic systems relevant for technological applications. 
Historically, the foundations for a description of thermal fluctuations in micromagnetic systems were laid by Brown when he investigated the thermal switching of single-domain particles\cite{BRO-59, BRO-63}. Today, these particles are used in promising biomedical applications such as disease detection and tumor treatment\cite{PAN-03,PAN-09, GAO-14, WU-15}. 
In order for these applications to be successful, a full understanding of the particles' thermal switching is important. For example, many characterization procedures \cite{BOG-15,Ludwig2016,ludwig2017analysis,LEL-15b, LEL-17,Leliaert2017Interpreting,Liebl2015} require this knowledge to accurately determine the particles' properties. Also diagnostic particle imaging \cite{WIE-12,baumgarten2015plane,COE-12, COE-15a, COE-17,ficko2015extended,Ficko2016,THE-17, GLE-05}, and therapeutic applications\cite{HER-06, PER-15} rely on models of the particles' thermal switching. Currently, these models are often based on approximations that not always take into account that, e.g. the magnetization state in large particles can deviate from a uniform magnetization\cite{KIM-15b}, or the particles might interact with each other via the magnetostatic interaction\cite{BRA-13}. In such cases, the analytical models do not accurately reflect the true magnetization dynamics of the particles, and one has to rely on numerical models, the most accurate of which are based on a micromagnetic approach \cite{LEL-14c,LAS-15,LEL-15,SHA-15,REE-17,ILG-17,CAB-17}. \\

Next to their relevance in magnetic nanoparticle research, thermal fluctuations also play an important role in (exchange-coupled) continuous magnetic systems.
One technologically relevant example is domain wall motion through a magnetic nanostrip, proposed as the operating principle for the racetrack memory\cite{PAR-08,HAY-08,PAR-15} and for logic devices\cite{ALL-05,ATK-06,BAR-06,VAN-15,OMA-14}.
Recently, even smaller magnetization structures, i.e. skyrmions, have been proposed in both memory\cite{FER-13} and logic devices\cite{ZHA-15a}. As the information carriers in these devices become smaller, the influence of thermal fluctuations further grows in importance: at such small spatial scales, the thermal stability of the bits themselves starts to become a relevant research question\cite{COR-17}. At the same time, their thermal depinning becomes an inherently stochastic process\cite{WUT-12}. When numerically investigating domain wall motion at low driving forces, the dynamics can only be captured by considering the interplay between thermal fluctuations, the disorder energy landscape of the material and the driving forces. The resulting motion is then called domain wall creep\cite{MET-07}. Until now, full micromagnetic simulations are still prohibitively expensive in all but the smallest of such systems\cite{LEL-16}.

Thermal fluctuations are also critical to the design of magnetic storage elements. They do not only determine the data retention limit of any magnetic storage system, but can also influence the read and write process of a MRAM cell. So estimation of read and write errors requires stochastic micromagnetic modeling of the spin valve during the application of the spin-torque current. This very challenging because of large time scales that are involved\cite{ROY-16}. 

There exist different theoretical approaches, each with their respective advantages and disadvantages, to study thermally induced magnetization dynamics\cite{REE-17,COF-12}. 
Following Brown\cite{BRO-63}, it is possible to derive the Fokker-Planck equation describing the time-dependent probability distribution of the magnetization directions of an ensemble of uniformly magnetized magnetic nanoparticles\cite{BRO-63,REE-17}. However, only in the simplest cases, e.g. when the particles' anisotropy axes are aligned with the applied field, an analytical solution can be found. In more complex cases approximations have to be introduced and when considering continuous systems consisting of several exchange-coupled finite difference cells, this approach becomes intractable.

Alternatively, thermal fluctuations can be included as a stochastic term in the Landau-Lifshitz-Gilbert (LLG) equation, henceforth named stochastic LLG (sLLG) equation by adding a thermal field term to the effective field. This approach was presented by Lyberatos\cite{LYB-93} and is based on the fact that finite difference cells can be considered as dipoles comparable to single-domain particles. This method has the advantage that it is a straightforward extension of the LLG equation. On the other hand, in contrast to the Fokker-Planck approach, the resulting sLLG equation has to be solved many times in order to gather enough data to draw conclusions about averaged quantities. 


Integrating stochastic differential equations (SDE) requires the use of non-Riemann calculus to be able to deal with the discontinuous thermal field term. A full discussion of this topic lies beyond the scope of this article, and for an excellent introduction we refer to Ref.~\cite{REE-17}. In general, SDE's are not trivial to numerically integrate, and require specialized methods\cite{KLO-12}, which are only suited to integrate SDE's written in either their Ito or Stratonovich form, as otherwise a drift term might appear in the solution. When considering variable time stepping algorithms, the complexity further increases\cite{MAU-98,BUR-03}, sometimes even canceling the relative advantage obtained by using such methods. 
However, Berkov and Gorn have shown that the drift term in the sLLG equation manifests itself only in the length of the magnetization vector\cite{BER-02a} which, in contrast to the Landau-Lifshitz-Bloch equation\cite{ATX-16}, is held constant. As shown in Ref.~\cite{BER-02a}, this can be seen more clearly when writing the sLLG equation in spherical coordinates. In Cartesian coordinates, numerical noise would build up in this direction if it weren't accounted for by  renormalizing the magnetization after each time step to ensure that the drift term does not influence the magnetization dynamics. Consequently, the Ito and Stratonovich interpretation are equivalent for integrating the sLLG equation, enabling the use of higher order solvers to integrate the sLLG eqation\cite{BER-05}. Despite this result, in literature one often still finds the recommendation to use the second order Heun's solver (here denoted with ``RK12'', because it is a second order Runge-Kutta type solver with embedded first order solution) with a very small fixed time step of the order of femtoseconds to simulate micromagnetic systems at finite temperatures\cite{LOP-12,GAR-98,VAN-14a}. In this paper, we present the use of higher order solvers with adaptive time stepping for such simulations and show that this method offers significant advantages.


\section{Methods}
\label{sec:method}
The Landau-Lifshitz-Gilbert (LLG) equation\cite{LAN-35, GIL-04} contains a precession and a damping term, and describes the magnetization dynamics at the nanometer length scale and the picosecond timescale.
\begin{equation}
	\dot{\mathbf{m}} =  -\frac{\gamma}{1+\alpha^2} \left(  \m \times \Beff  +\alpha \m \times \left( \m \times \Beff \right)   \right) \label{eq:LLG}
\end{equation}
In this equation, $\gamma$ denotes the gyromagnetic ratio, $\alpha$ the dimensionless Gilbert damping parameter and \Beff the effective field.

There is more than one way to add thermal fluctuations to the LLG equation. We choose to add a stochastic thermal field $B_\mathrm{therm}$ as a contribution to the effective field term in both the precession and damping term, although it has been shown that the field contribution could also be omitted in the damping term if the size of the thermal field is rescaled adequately\cite{GAR-98}. A second common option\cite{REE-17} is to add thermal fluctuation directly as an extra thermal torque term to the LLG equation.

The properties of the thermal field $B_\mathrm{therm}$ were determined by Brown when he investigated the thermal switching of single-domain particles\cite{BRO-63}. Later, it was realized that this theory was also applicable to micromagnetic simulations as each finite difference cell can be considered as such a particle \cite{LYB-93}. The thermal field is given by 
\begin{eqnarray}
	\label{eq:th1}
	\av{\mathbf{B}_\mathrm{therm}}=0\\
	\label{eq:th2}
	\av{\mathbf{B}_{\mathrm{therm},i}(t)\mathbf{B}_{\mathrm{therm},j}(t')}=q\delta(t-t')\delta_{ij}\\
	\label{eq:th3}
	q=\frac{2k_\mathrm{B}T\alpha}{M_\mathrm{s}\gamma V}
\end{eqnarray}
Here, the operator $\av{\cdot}$ denotes a time average, $\av{\cdot\cdot}$ a correlation, $\delta$ the Dirac delta function and the indices $i$ and $j$ run over the $x$, $y$ and $z$ axes in a Cartesian coordinate system. The thermal field has zero average [Eq. (\ref{eq:th1})], is uncorrelated in time and space [Eq. (\ref{eq:th2})] and its size $q$ is given by Eq. (\ref{eq:th3}). In this equation, $k_\mathrm{B}$ denotes the Boltzmann constant, $T$ the temperature, $M_\mathrm{s}$ the saturation magnetization, and $V$ the volume on which the thermal fluctuations act, i.e. the volume of a single finite difference cell.\\
Equations (\ref{eq:th1}) to (\ref{eq:th3}) are determined such that the effect of the thermal fluctuations is independent of the spatial discretization used: when splitting up a volume into subvolumes and averaging the thermal fluctuations within those, one will recover the same resulting dynamics as in the undivided volume. The same is also true for the time step $\Delta t$: when averaged out over a larger time, thermal fluctuations decrease in strength and again, this proportionality is determined such that the average dynamics do not depend on the time discretization. 

\subsection{Implementation in \mumax}
\mumax is a GPU-accelerated micromagnetic software package which numerically solves the LLG equation using a finite-difference discretization\cite{VAN-14a}. The thermal field is included in the effective field as 
\begin{equation}
	\label{eq:thermalfield}
	\mathbf{B}_\mathrm{therm} = \boldsymbol\eta \sqrt{ \frac{2\alpha k_\mathrm{B} T}{M_\mathrm{s}\gamma V\Delta t} }
\end{equation}
where $\Delta t$ denotes the time step and $\boldsymbol\eta$ is a random vector drawn from a standard normal distribution whose value is redetermined after every time step\cite{LOP-12,VAN-14a}. 

\mumax provides several explicit Runge-Kutta methods to time step the LLG equation, the details of which can be found in Ref.~\cite{VAN-14a}. Here, we will only mention the ones relevant for this work. Previously, simulations at nonzero temperatures in \mumax were performed with the widely used Heun's method (RK12), using a very small time step of the order of 5 fs. 
In Section \ref{sec:validation}, we will validate our results by comparing them to the solution obtained with this solver when there are no analytical solutions available.

For dynamical simulations at zero temperature, the default solver is the Dormand-Prince method (RK45). This solver offers 5th order error convergence and contains an embedded 4th order method to estimate the error. Generally speaking (i.e. when not investigating very fast dynamics, or in the absence of thermal fluctuations), it is not advantageous to implement even higher-order solvers. The reason for this is threefold: 1) For the moderately small torques encountered in typical micromagnetic simulations, the performance of the solver is limited by its stability regime. This means that using even slightly larger time steps will result in much larger errors no matter how small the exerted torques are. 2) Higher order methods typically need more intermediate torque evaluations per time step, thus reducing the advantage obtained by taking a larger time step. 3) Due to the memory required to store the results of the intermediate torque evaluations, higher-order solvers disproportionately increase the memory consumption compared to the obtained gain in performance.

\subsection{Sixth order Runge-Kutta-Fehlberg solver}
Due to the large size of the stochastic thermal field, simulations at nonzero temperatures require much smaller time steps, so that the solver performance is not longer limited by its stability regime, and large enough time steps can be taken to justify the additional intermediate torque evaluations. Therefore, we also implemented the 6th order Runge-Kutta-Fehlberg (RKF56) method with 5th order embedded solution shown in Table~\ref{tab:rk56}.
\begin{table}[ht!]
\begin{tabular}{c|cccccccc}
$0$&&&&&&&&\\
$\frac{1}{6}$&$\frac{1}{6}$&&&&&&&\\
$\frac{4}{15}$&$\frac{4}{75}$&$\frac{16}{75}$&&&&&&\\
$\frac{2}{3}$&$\frac{5}{6}$&$-\frac{8}{3}$&$\frac{5}{2}$&&&&&\\
$\frac{4}{5}$&$-\frac{8}{5}$&$\frac{144}{25}$&$-4$&$\frac{16}{25}$&&&&\\
$1$&$\frac{361}{320}$&$\frac{-18}{5}$&$\frac{407}{128}$&$-\frac{11}{80}$&$\frac{55}{128}$&&&\\
$0$&$-\frac{11}{640}$&$0$&$\frac{11}{256}$&$-\frac{11}{160}$&$\frac{11}{256}$&$0$&&\\
$1$&$\frac{93}{640}$&$-\frac{18}{5}$&$\frac{803}{256}$&$-\frac{11}{160}$&$\frac{99}{256}$&$0$&$1$&\\\hline
&$\frac{31}{384}$&$0$&$\frac{1125}{2816}$&$\frac{9}{32}$&$\frac{125}{768}$&$\frac{5}{66}$&$0$&$0$\\
&$\frac{7}{1408}$&$0$&$\frac{1125}{2816}$&$\frac{9}{32}$&$\frac{125}{768}$&$0$&$\frac{5}{66}$ &$\frac{5}{66}$\\\hline
&$-\frac{5}{66}$&$0$&$0$&$0$&$0$&$-\frac{5}{66}$&$\frac{5}{66}$&$\frac{5}{66}$
\end{tabular}
\caption{Butcher tableau\cite{BUT-08} of the Runge-Kutta-Fehlberg (RKF56) solver with sixth order solution and embedded 5th order solution. The difference between both, used as error estimate, is given in the last row.}
\label{tab:rk56}
\end{table}

Unlike the RKF56 solver, some of the solvers used in \mumax like the RK45 method, benefit from the first-same-as-last (FSAL) property. In these solvers, the last torque evaluation of the current time step corresponds to the first evaluation of the next time step, thus effectively reducing the number of evaluations per step by 1. However, as the stochastic thermal field is not constant in between time steps, the torque continuity requirement is not longer fulfilled, and the first and last torque evaluations have to be performed separately. Because the RKF56 solver never has the FSAL property, its performance can only compete with these other methods at nonzero temperatures, where the other methods do not benefit from the FSAL property either.

\begin{figure}[ht]
	\includegraphics[width=8.6cm]{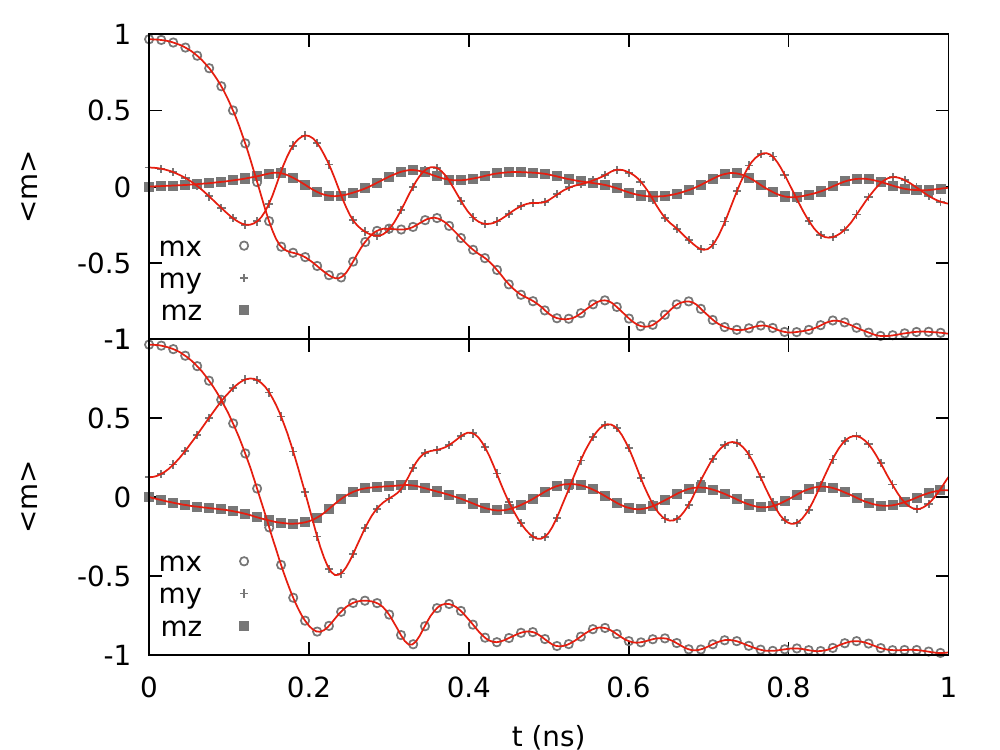}\\
	\includegraphics[width=3.1cm]{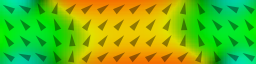}
	\includegraphics[width=3.1cm]{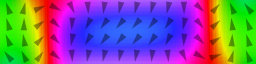}
\caption{\label{fig:std4} The solutions to standard problem 4, obtained by \mumax (gray points) and OOMMF (red lines). This problem focuses on micromagnetic dynamics and looks at the time evolution of the magnetization during the relaxation of a magnetic rectangle from an initial s-state. The problem is run for two different applied fields (top and bottom graph) and the space dependent magnetizations when $\langle m_\mathrm{x}\rangle$ crosses zero are shown below, in the left and right plot, respectively.}
\end{figure}

The implementation of the Sixth order Runge-Kutta-Fehlberg solver is subject to the same tests used for the other solvers implemented in \mumax\cite{VAN-14a}, and a full report is beyond the scope of this article. Nonetheless, Fig.~\ref{fig:std4} shows that our solution to standard problem 4, proposed by the $\mu$Mag modeling group\cite{MAG}, agrees with the solution obtained with OOMMF\cite{DON-99} (not using the RK56 solver).

\begin{figure}[ht]
	\includegraphics[width=8.6cm]{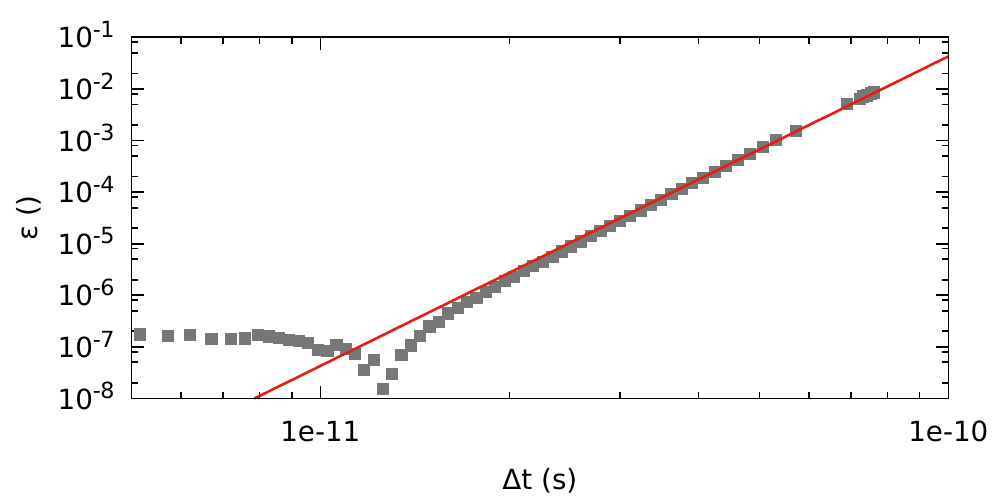}
	\caption{\label{fig:convergence} The error $\epsilon$ as function the time step $\Delta t$ of the RKF56 solver (gray squares) displays the sixth order convergence (red line).}
\end{figure}

One of the most sensitive checks one can perform to test the implementation of a solver is to investigate the scaling of its error convergence. Figure~\ref{fig:convergence} shows the error $\epsilon$ after a single precession without damping of a single spin in a field of 0.1 T as function of the time step $\Delta t$. The solver shows the expected sixth order convergence up to the limit of the single precision implementation\cite{VAN-14a} ($\epsilon\approx 10^{-7}$).

\subsection{Time stepping with adaptive time steps}
When performing simulations at nonzero temperatures, it is important to note that the size of the thermal field is determined by 1/$\sqrt{\Delta t}$. This implies that, when a large thermal field is generated leading to a bad step (defined as a step where the torque was too large for the used time step), the step will be undone, and the adaptive time step algorithm will decrease the time step, thus further increasing the size of the field. Luckily, this $1/\sqrt{\Delta t}$ dependency makes the time step smaller at a slower rate than that the error is reduced ($\Delta t^{N}$ with $N$ the order of the solver\cite{GUS-92}). However, it is important to use higher order solvers like the RK45 or RKF56 method in order to maintain a large time step.

To give the correct solution, the statistical properties of the random numbers $\boldsymbol\eta$ in Eq.(\ref{eq:thermalfield}) should correspond to the ones determined in Eqs. (\ref{eq:th1}) to (\ref{eq:th3}). However, if one would redraw these random numbers after a bad step, the small thermal fields (small $\eta$) would be applied during longer time steps and large thermal fields (large $\eta$) during shorter time steps, thus virtually changing the distribution of the random numbers, and eventually giving rise to incorrect solutions. In our implementation we avoid this by keeping the previously drawn random numbers and rescaling the thermal field with a factor 
$\sqrt{\Delta t_\mathrm{new}/\Delta t_\mathrm{bad step}}$
in case a bad step is encountered to ensure that the correct statistical properties of the thermal field are maintained.

\section{Validation}
\label{sec:validation}
The adaptive time stepping at nonzero temperatures will be tested in several test cases, focusing on different aspects, i.e. static vs. dynamic properties of uncoupled spins or continuous magnets. Each time, the simulation results obtained with adaptive time stepping will be compared either to analytical solutions or to the solutions obtained with the RK12 method with fixed time step. 
 
\subsection{Spectra of a single spin}
\begin{figure}[ht]
	\includegraphics[width=8.6cm]{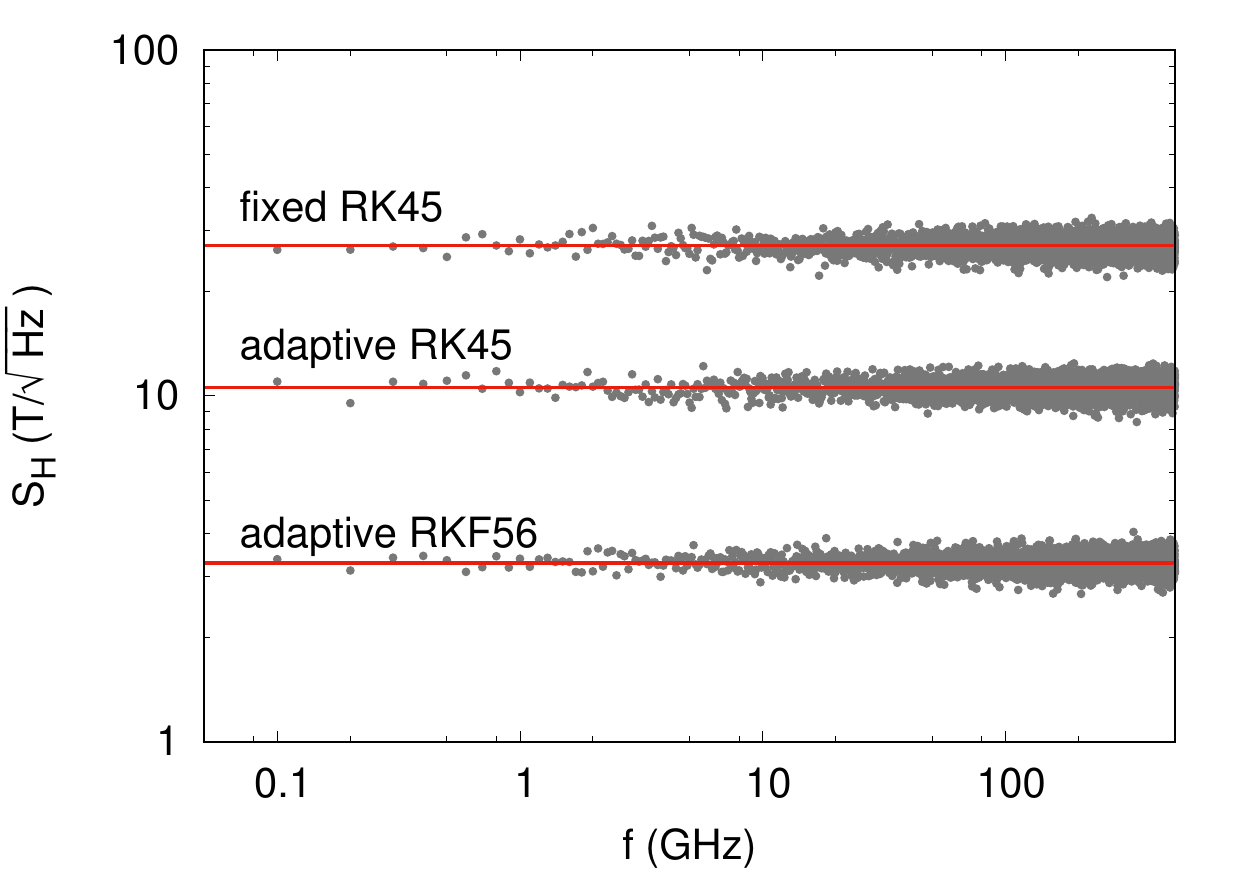}
	\caption{\label{fig:fieldspectrum} The thermal field spectra of a single spin (rescaled with an arbitrary factor for clarity reasons) obtained with the RK45 solver with fixed time steps, and with the RK45 and RKF56 solver with adaptive time steps. All spectra display white noise.}
\end{figure}

It will be verified whether the thermal field and the resulting magnetization dynamics of a single spin in the absence of an external field shows the theoretically expected behavior. The thermal field should display a white spectrum $S_\mathrm{H}$ construction, as its size is given by Gaussian random numbers. Figure~\ref{fig:fieldspectrum} proves that this is indeed the case.

\begin{figure}[ht]
	\includegraphics[width=8.6cm]{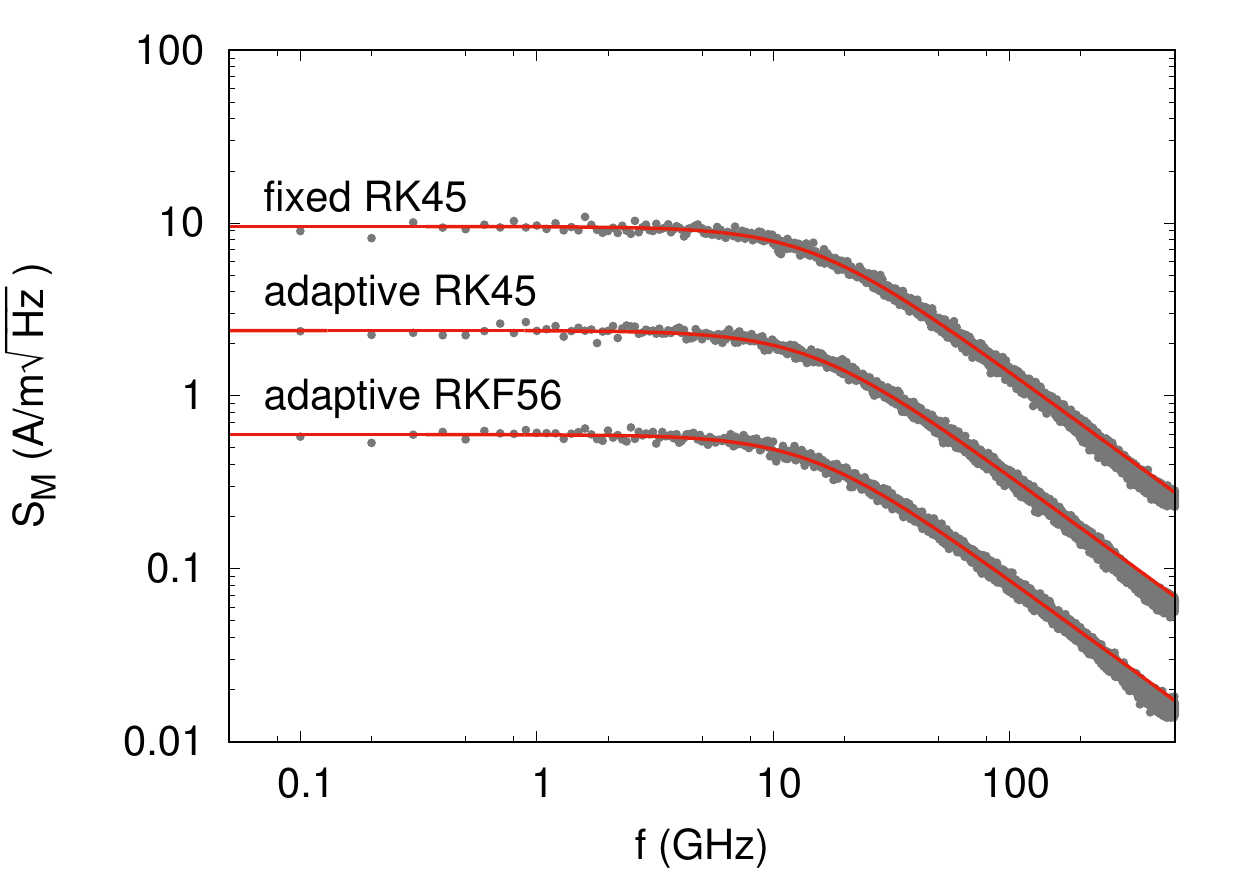}
	\caption{\label{fig:magspectrum} The magnetization spectra of a single spin (rescaled with an arbitrary factor for clarity reasons) obtained with the RK45 solver with fixed time steps, and with the RK45 and RKF56 solver with adaptive time steps. All spectra display the theoretically expected shape depicted by the red lines.}
\end{figure}

The random thermal fields acting on a single isotropic finite-difference cell gives rise to a random walk on the unit sphere\cite{REE-17,LEL-15b}. The shape of the spectral density $S_\mathrm{M}(f)$ of such a random walk is described by the square root of a Lorentzian\cite{MAC-54},
\begin{equation}
\label{eq:spectral}
S_\mathrm{M}(f)\sim\sqrt{\left(\frac{f_0/2}{f_0^2+(\pi f)^2}\right)},
\end{equation}
i.e. white noise with a 1/$f$ cutoff at a cutoff frequency $f_0$ given by\cite{GAR-98,ILG-17}
\begin{equation}
\label{eq:cutoff}
f_0=\frac{\alpha}{(1+\alpha^2)}\frac{\gamma k_\mathrm{B}T}{M_\mathrm{s}V}
\end{equation}
Figure~\ref{fig:magspectrum} shows the obtained magnetization spectra (gray dots) indeed coincide with the red lines determined by Eqs.~(\ref{eq:spectral}) and (\ref{eq:cutoff}). Because all spectra coincide, they were rescaled with an arbitrary factor for clarity.

\subsection{Equilibrium magnetization}
\begin{figure}[ht]
	\includegraphics[width=8.6cm]{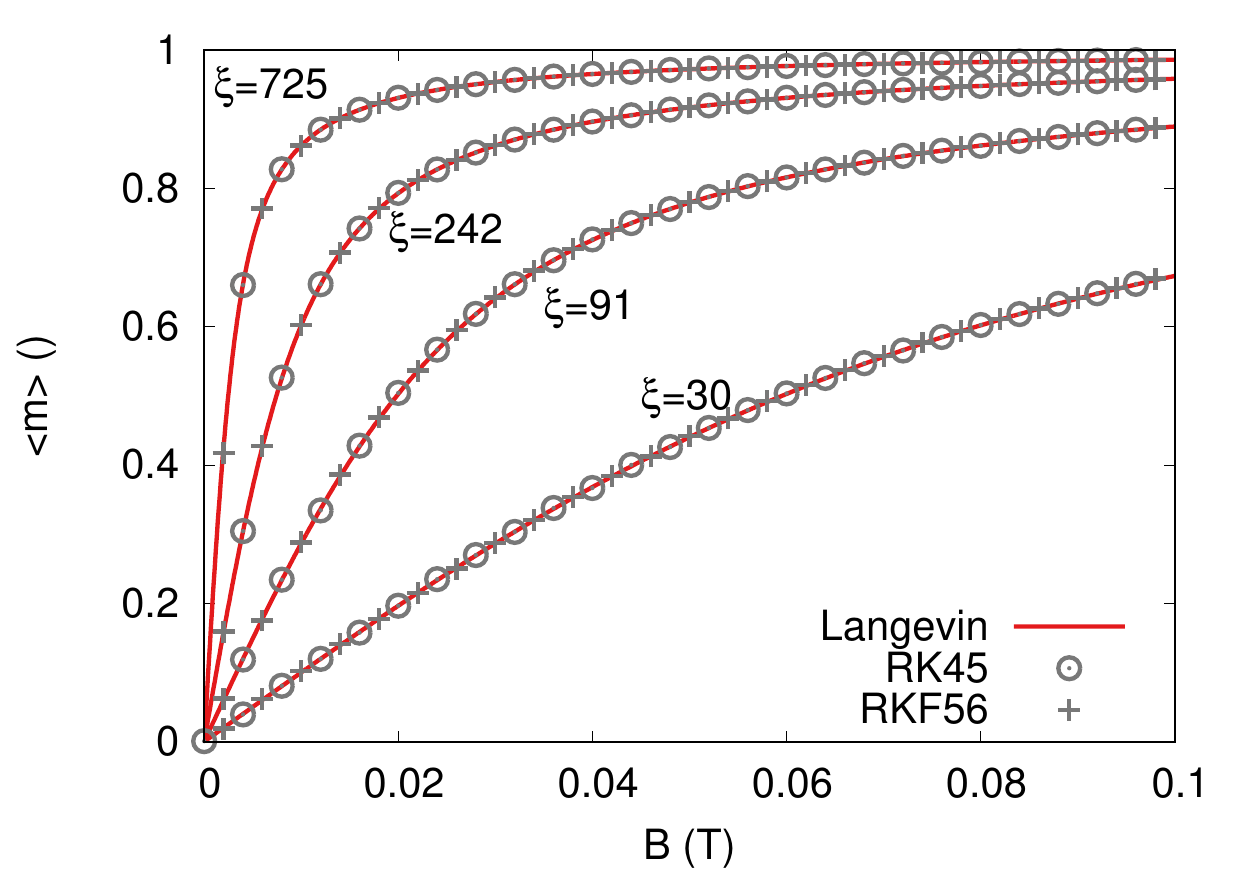}
	\caption{\label{fig:langevin} The average magnetization $\langle m\rangle$ of an ensemble of $2^{18}$ uncoupled spins in thermal equilibrium at different temperatures and different cell sizes (reflected in the different values of $\xi$) for the RK45 (gray circles) and RKF56 (gray crosses) with adaptive time steps. The results agree perfectly with the Langevin function [Eq.~(\ref{eq:langevin})], shown by a red line.}
\end{figure}

This validation problem checks whether the magnetization of an ensemble of uncoupled spins in thermal equilibrium in an externally applied field is described by the Langevin function $\mathcal{L}(\xi)$,
\begin{equation}
	\label{eq:langevin}
	\mathcal{L}(\xi)=\coth(\xi)-\frac{1}{\xi}
\end{equation}
where the argument $\xi$ stands for
\begin{equation}
	\xi=\frac{\mu_0M_{\mathrm{s}} V H_{\mathrm{ext}}}{k_\mathrm{B} T}.
	\label{eq:langevin_argument}
\end{equation}
Figure~\ref{fig:langevin} proves that this is indeed the case for 4 different $\xi$ (realized by 2 different cell sizes and 2 different temperatures) simulated with the RK45 and RKF56 solver with adaptive time steps over a large range of applied fields. 

\subsection{Thermal switching}
\label{sec:bench1}
After the equilibrium magnetization addressed in the previous problem, we will now concern ourselves with a dynamical problem consisting of the thermal switching rate of a single (macro-)spin particle with uniaxial anisotropy. In the limit of a high energy barrier (compared to the thermal energy), the switching rate $\nu$ is given by\cite{BRE-12}
\begin{equation}
\nu = \gamma\frac{\alpha}{1+\alpha^2}\sqrt{\frac{8 K^3 V}{2\pi M_\mathrm{s}^2k_\mathrm{B}}}e^{-KV/k_\mathrm{B}T}. 
\label{eq:switch}
\end{equation}

For an ensemble of uncoupled spins, initialized with all spins pointing in the same direction, this switching gives rise to an exponentially decaying magnetization, with a decay constant $1/2\nu$. In this test problem, we simulate this decay to determine the numerical switching rate, and compare these values to their theoretical prediction.

\begin{figure}[ht]
	\includegraphics[width=8.6cm]{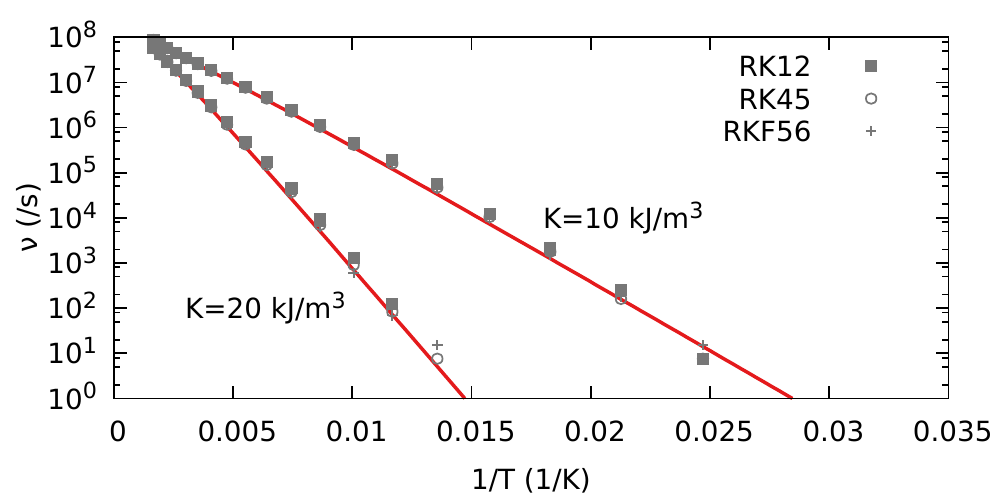}
	\caption{\label{fig:arrhenius} Arrhenius plot of the thermal switching rate of a 10\,nm  large cubic cell, with $M_\mathrm{s}$=1 MA/m, $\alpha$=0.1,$K$=10 or 20 kJ/m$^3$. Simulations were performed using the RK12 solver with fixed time steps ($\Delta t$=5 fs) or the RK45 or RKF56 solver with adaptive time steps on an ensemble of $2^{18}$ non-interacting cells for 1\,$\mu$s  or until the ensemble magnetization crossed 0. All results agree with the red solid lines depicting the analytically expected switching rates (Eq.~\ref{eq:switch}).}
\end{figure}

Figure~\ref{fig:arrhenius} shows Arrhenius plots for the temperature-dependent switching rate $\nu$ of uncoupled finite difference cells with volume $V$=(10\,nm)$^3$ and uniaxial anisotropy constant $K$=1\E{4} or 2\E{4} J/m$^3$. Again, a quantitative agreement is seen between the \mumax simulations and the theoretically predicted behavior.

\subsection{Thermally excited magnetization spectrum}
\label{sec:bench2}
\begin{figure}[ht]
	\includegraphics[width=8.6cm]{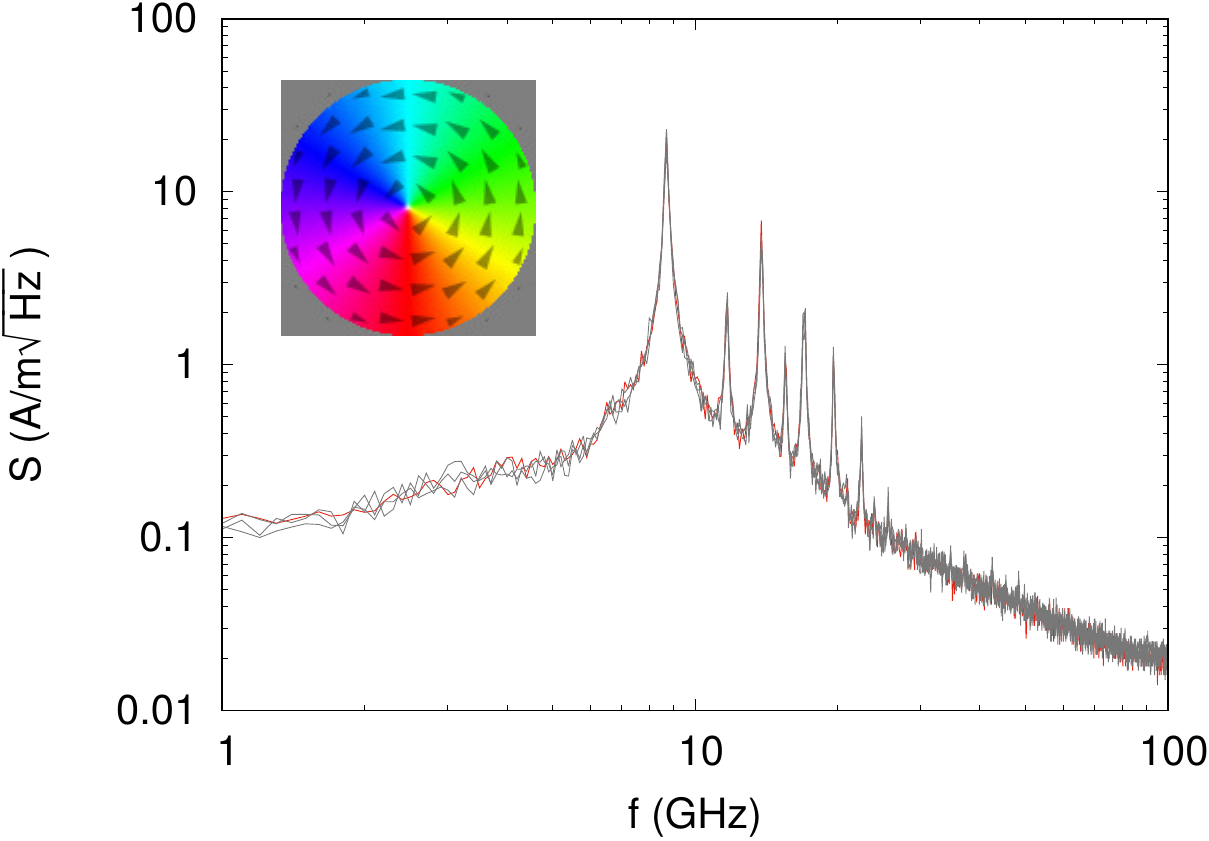}
	\caption{\label{fig:circle} The red line corresponds to the spectrum obtained with the Heun solver with $\Delta t$=5 fs. The three gray lines, which overlap almost perfectly with the red one, correspond to the RK45 solver with fixed time step $\Delta t$=300 fs, and with the solution obtained with the RK45 and RKF56 solver with adaptive time stepping with $\epsilon=10^{-5}$. All spectra were averaged out over 25 realizations with a different random seed for the thermal field. The inset shows the equilibrium vortex magnetization structure in the system under consideration\footnote{The minimum in the error is a result between a direct match between the used time step and the gyration period used for the evaluation of the precision\cite{VAN-14a}.}}
\end{figure}

In this problem we look at the thermally excited magnetization spectrum of a 10 nm thick disk with a diameter of 512 nm, $M_\mathrm{s}$=1 MA/m, exchange constant $A_\mathrm{ex}$=10 pJ/m, and $\alpha=1$, discretized in cells measuring 4 by 4 by 10 nm$^3$. The equilibrium magnetization structure in such a disk is a vortex structure, as depicted in the inset of Fig.~\ref{fig:circle}. We apply a thermal field corresponding to 300 K, thereby thermally exciting the sample, resulting in the spectra shown in Fig.~\ref{fig:circle}. The spectrum depicted in red was obtained with the RK12 solver with fixed time step ($\Delta t$=5 fs), and serves as a reference solution. The gray lines, which agree almost perfectly with the benchmark solution, correspond to the three spectra obtained with the RK45 solver with the time step fixed to $\Delta t$=300 fs and with the RK45 and RKF56s solvers with adaptive time step and $\epsilon=10^{-5}$. Note that the spectra overlap even at high frequencies, indicating that the adaptive time stepping does not lead to spectral leakages.

\subsection{Thermal diffusion of a domain wall}
\label{sec:bench3}
\begin{figure}[ht]
	\includegraphics[width=6.4cm]{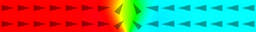}
	\caption{\label{fig:domainwall} The transverse domain wall in the center of the nanowire. By compensating the edge charges, we simulate an infinitely long wire.
    }
\end{figure}

In a last validation problem we investigate the thermal driven diffusion of transverse domain walls in a non-disordered permalloy nanowire\cite{LEL-15a}. We simulate a nanowire with cross-sectional dimensions of $100\times 10$ nm$^2$ discretized in cells of $3.125\times 3.125\times 10$nm$^3$. We use the material parameters of permalloy: $M_\mathrm{s}=860$ kA/m, $A\mathrm{ex}=13$ pJ/m, $\alpha$=0.01 and simulate the domain wall in the center of a moving window, as shown in Fig.~\ref{fig:domainwall}. 

The thermally driven motion can be described by a random walk characterized by a diffusion constant $D$ which scales linearly with temperature. Similarly as in Ref. \cite{LEL-15a} we simulate a transverse domain wall for 100 ns and repeat this simulation with a large number of different random seeds. In Tab.~\ref{tab:diffusion}, we compare the results obtained from the full micromagnetic simulations with the diffusion constant $D$ of approximately 310 nm$^2$/ns predicted by the model introduced, and numerically validated in Ref.~\cite{LEL-15a}using the RK12 solver with fixed time step. 
The results show that the standard errors $s$ are larger than the difference between the obtained diffusion constants and the expected value. This also indicates that, for this problem, an error tolerance $\epsilon=10^{-3}$ suffices for all practical purposes, as the variance between different simulations gives rise to an uncertainty that is larger than the errors due to the use of this relatively large $\epsilon$. As this is problem dependent, the default value in \mumax remains $10^{-5}$, but we suggest that, depending on the system under consideration, larger values might be suitable in simulations at nonzero temperatures.

\begin{table}[!ht]
\caption{The diffusion constant $D$ and standard error $s$, determined from simulations performed using several solvers, with adaptive time stepping with error tolerance $\epsilon$ and repeated for a total number of $N$ realizations at 300 K. The theoretically predicted $D$ approximately equals 310 nm$^2$/ns.}
\begin{tabular}{l|c|c|c|c}
Solver&$\epsilon$&$N$&$D$&$s(D)$\\
&()&()&(nm$^2$/ns)&(nm$^2$/ns)\\
\hline
RK23&$5\times10^{-3}$&500&315&19\\
RK45&$5\times10^{-3}$&500&309&19\\
RK45&$1\times10^{-3}$&1000&316&14\\
RKF56&$5\times10^{-3}$&200&340&37\\ 
RKF56&$1\times10^{-3}$&200&335&32\\ 
RKF56&$1\times10^{-4}$&200&315&31
\end{tabular}
\label{tab:diffusion}
\end{table}
\section{Performance}
\label{sec:performance}
\begin{figure}[ht]
	\includegraphics[width=8.6cm]{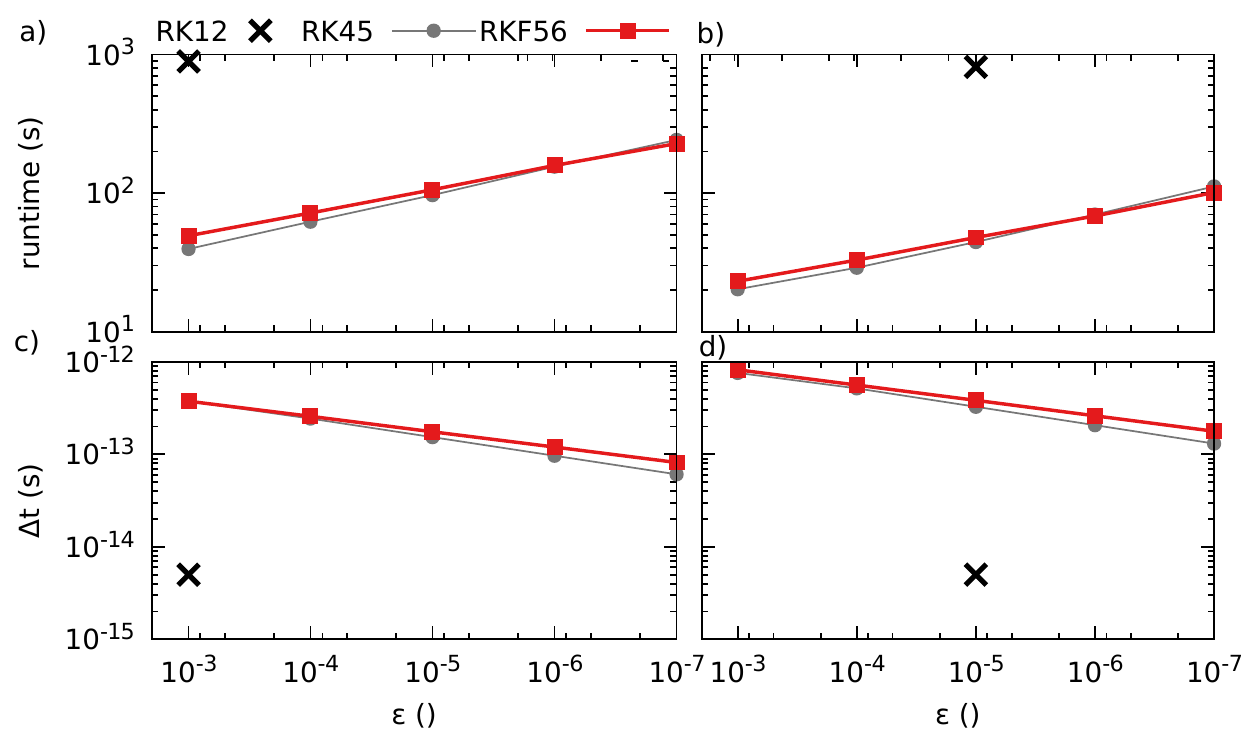}\\
\caption{\label{fig:bench} Benchmark results for the systems described in Sections~\ref{sec:bench2} [panel a) and c)] and \ref{sec:bench3} [b) and d)], respectively. The top row shows the runtime required to simulate 10 ns of magnetization dynamics, while the bottom row shows the average time step used. The black crosses indicate the performance of the fixed time step RK12 method with $\epsilon$ estimated from the difference between the second order and embedded first order solution. All results were obtained an NVIDIA Titan Xp GPU in a system running on a 7th generation i5 CPU.}
\end{figure}

To assess the performance of the presented methods we will consider the problems detailed in \ref{sec:bench2} and \ref{sec:bench3}, i.e the thermal spectrum of a disk and thermally driven domain wall diffusion, respectively, as benchmarks. The benchmark results are shown in Fig.~\ref{fig:bench} a) and b). For each of these problems, the simulation time was first determined when solving the problem with the RK12 solver with a fixed time step of 5 fs. As this is a second order solver with embedded first order solver, the difference between both solutions serves as an error estimate, allowing us to estimate with which error tolerance $\epsilon$ in the adaptive time stepping the results should be compared. This data point is indicated by a black cross in the figure. Next, the problems were solved using adaptive time stepping, once with the RK45, and once with the RKF56 solver, with $\epsilon$ ranging from $10^{-3}$ to $10^{-7}$, shown in gray and red, respectively.
The simulation runtime [Figs.~\ref{fig:bench} a) and b)] show the time it took to simulate 10 ns of magnetization dynamics, while Figs.~\ref{fig:bench} c) and d) indicate the average time step $\Delta t$ used by the solver, for each $\epsilon$. 
When comparing the results from the adaptive time stepping methods with the result of the RK12 solver at the same estimated $\epsilon$, one sees that the adaptive time stepping methods use a considerably larger time step without a loss of accuracy. For these two benchmark problems, this results in a twenty fold speedup of the simulation. 

We also investigated the performance of the system described in Section~\ref{sec:bench1}, i.e. thermal switching of uncoupled spins. There, an even higher speedup was achieved, but this was attributed to the fact that such systems do not require the calculation of the demagnetizing field so that other factors, like the generation of the random numbers for the thermal field, become the limiting factor. Because the random numbers have to be generated only once per time step independently of the used solver, a very large performance gain can be achieved by using higher order solvers which allow time steps that are over a 1000 times larger than the ones necessary for the RK12 method with the same accuracy. However, as this highly depends on the used hardware, the performance gained by using the adaptive time stepping methods can lie anywhere between a factor 20 to 10 000, depending on accuracy.  These observations indicate that the presented methods are particularly suited for magnetic nanoparticle research, where micromagnetic simulations are becoming increasingly important\cite{CAB-17}. 

The relative performance of the RK45 and RKF56 solver is comparable and show the expected trends that the RK45 solver is faster at higher $\epsilon$ and simulated temperatures while the RKF56 solver is faster at low $\epsilon$ and temperatures. Generally, only when simulating systems at very high temperatures, or with very small $\epsilon$, it does pay off to use the RKF56 solver.

\section{Conclusions}
\label{conclusion}
In this paper, we have exploited the fact that the drift term in the stochastic Landau-Lifshitz-Gilbert equation is only able to manifests itself in the direction of the magnetization length, which is fixed. Therefore, we were able to straightforwardly extend existing high order solvers with adaptive time stepping at nonzero temperatures. In an effort to further increase the performance, we have implemented the sixth order Runge-Kutta-Fehlberg solver, and we extensively validated both the correctness of this newly implemented solver and the adaptive time stepping method used at nonzero temperature. All presented methods are included in the open-source micromagnetic software package \mumax and are thus freely available online. 

The main advantages of the presented adaptive time stepping methods at nonzero temperatures are that they offer an inherent error control, which is unavailable with fixed time stepping methods, and without a loss of accuracy one can obtain a twenty fold speedup compared to the commonly best practice of using the RK12 solver with small fixed time step. This enables simulations which previously took too long to be considered feasible and will be useful for micromagnetic research of continuous (exchange coupled) systems like spin valves, or domain wall motion in nanowires, and for uncoupled spins, e.g. in magnetic nanoparticle research.

\section{Acknowledgement}
This work was supported by the Fonds Wetenschappelijk Onderzoek (FWO-Vlaanderen) through Project No. G098917N and a postdoctoral fellowship (A.C.).
J. L. is supported by the Ghent University Special Research Fund (BOF postdoctoral fellowship). We gratefully acknowledge the support of NVIDIA Corporation with the donation of the Titan Xp GPU used for this research.

%
\end{document}